\documentclass[aps,showpacs,superscriptadress,preprint]{revtex4}
\usepackage{graphicx}

\begin{document}
\title{A consistent thermodynamic treatment for quark mass density-dependent model}

\author{Shaoyu Yin$^1$ and Ru-Keng
Su$^{1,2}$\footnote{rksu@fudan.ac.cn}} \affiliation{
\small 1. Department of Physics, Fudan University,Shanghai 200433, China\\
\small 2. CCAST(World Laboratory), P.O.Box 8730, Beijing 100080, China\\
}

\begin{abstract}
The ambiguities and inconsistencies in previous thermodynamic
treatments for the quark mass density-dependent model are addressed.
A new treatment is suggested to obtain the self-consistent results.
A new independent variable of effective mass is introduced to make
the traditional thermodynamic calculation with partial derivative
still practicable. The contribution from physical vacuum has been
discussed. We find that the properties of strange quark matter given
by quark mass density-dependent model are nearly the same as those
obtained by MIT bag model after considering the contribution of the
physical vacuum.
\end{abstract}


\maketitle

\section{Introduction}

It is generally accepted that the effective masses of particles will
change with density due to medium effects. Many theoretical
considerations, including the finite temperature quantum hadron
dynamics (QHD) model\cite{Walecka:1997}, Brown-Rho
scaling\cite{Brown:1991}, finite temperature quark-meson coupling
(QMC) model\cite{Saito:1997}, and \textit{etc.}, had been suggested
to investigate the effective masses of mesons and nucleons. Besides
theoretical study, many experimental results which predict the
changes of particle masses with density have been shown. In
particular, the experiments of TAGX collaboration had shown directly
that when the density of the nucleon medium equals to $0.7n_0$ where
$n_0$ is the saturation density, the effective mass of neutral
$\rho$-meson reduces to 610 MeV\cite{Lolos:1998}. Both theoretical
and experimental results confirm that the medium effects are
important for studying the nuclear or quark systems.

To illustrate the medium effects more transparently in theory,
instead of first principle calculation, many authors introduced
different hypothesis to represent the medium contributions, for
example, supposed the density-dependent vacuum energy $B(\rho)$ to
modify QMC model\cite{Jin:1996dual,Wang&Song:1999}, suggested the
density-dependent NN$\rho$ coupling to address liquid-gas phase
transition\cite{Qian:2000&2001}, and \textit{etc.}. Employing these
hypothesis, many physical properties of nuclear matter, quark
matter, nucleon system and hyperon system had been discussed. In
particular, to simplify the calculations, many ideal quasiparticle
models in which the effective mass depends on the density and/or the
temperature had been suggested for studying the quark gluon plasma
\cite{Bannur:dual,Bluhm:2005}, gluon plasma \cite{Yang:1995} and
strange quark matter \cite{Fowler:1981}.

The quark mass density-dependent (QMDD) model \cite{Fowler:1981} is
one of such candidates. It was first suggested by Fowler, Raha and
Weiner. In this model, the masses of $u$, $d$ quarks and $s$ quark
are given by
\begin{equation}
m^*_{q}=\frac{B}{3\rho_B},\qquad(q=u,\overline{u},d,\overline{d}),
\end{equation}
\begin{equation}
m^*_{s,\overline{s}}=m_{s0}+\frac{B}{3\rho_B},
\end{equation}
where $B$ is a constant and $m_{s0}$ is the current mass of the
strange quark; and $\rho_B$, the baryon density, is defined as
\begin{equation}
\rho_B=\frac{1}{3}(\Delta \rho_{u}+\Delta \rho_{d}+\Delta \rho_{s}),
\end{equation}
where
\begin{equation}
\Delta
\rho_{i}=\rho_{i}-\rho_{\overline{i}}=\frac{g_{i}}{(2\pi)^{3}}\int
d^{3}k \left(\frac{1}{\exp[\beta(\epsilon_{i}(k)-\mu_{i})]+1} -
\frac{1}{\exp[\beta(\epsilon_{i}(k)+\mu_{i})]+1}\right),
\end{equation}
where $g_i$ is the degeneracy, $i$ ($i=u,s,d$) and $\overline{i}$
correspond to the quarks and antiquarks, respectively. The
\textit{ans\"{a}tze} in Eqs.(1) and (2) corresponds to a quark
confinement mechanism because if quark goes to infinity, the volume
of the system tends to infinity and the baryon number density goes
to zero, then $m^*_q$ approaches to infinity according to Eqs.(1)
and (2). The infinite mass prevents the quark from going to
infinity.

Employing QMDD model and considering weak processes
\begin{equation}
u+d\leftrightarrow u+s,\qquad s\rightarrow
u+e^{-}+\overline{\nu}_{e},\qquad d\rightarrow
u+e^{-}+\overline{\nu}_{e},\qquad u+e^{-}\rightarrow d+\nu_{e},
\end{equation}
and the condition of charge neutrality
\begin{equation}
2\Delta \rho_{u}=\Delta \rho_{d}+\Delta \rho_{s}+3\Delta \rho_{e},
\end{equation}
many physical properties of strange quark matter
\cite{Chakrabarty:1989,Chakrabarty:1991,Chakrabarty:1993,
Benvenuto:1995a,Benvenuto:1995b,Peng:2000,Zhang:2001,Peng:1999,
Wen:2005,Wang:2000,Zhang:2002,Zhang:2003, Wu:2005,Mao:2006} and
strange quark star \cite{Gupta:2003,Shen:2005} have been studied,
and the results are in good agreement with those given by MIT bag
model \cite{Farhi:1984}.

Although the density-dependent quark masses Eqs.(1) and (2) can
mimic the quark confinement mechanism, when we discuss the
thermodynamic behaviors of the system with such quarks, many
difficulties will emerge. The dispersion relation for quark
\begin{equation}
\epsilon(k,\rho)=[k^2+m^*(\rho)^2]^{1/2}
\end{equation}
will make many usual thermodynamic relations with partial derivative
no longer satisfied. As is well known in thermodynamics, a proper
choice of independent variables will have a suitable characteristic
thermodynamical function, from which all the thermodynamic
quantities can be obtained by partial derivatives without
integration. For example, with variables temperature $T$, volume $V$
and chemical potential $\mu$, the characteristic function is the
thermodynamical potential $\Omega=\Omega(T,V,\mu)$. From the
differential relation for reversible process
\begin{equation}
d\Omega=-SdT-pdV-\overline{N}d\mu,
\end{equation}
we have
\begin{equation}
S=-\left(\frac{\partial \Omega}{\partial T}\right)_{V,\mu},\qquad
p=-\left(\frac{\partial \Omega}{\partial V}\right)_{T,\mu},\qquad
\overline{N}=-\left(\frac{\partial \Omega}{\partial
\mu}\right)_{T,V},
\end{equation}
where $S$, $p$ and $\overline{N}$ are entropy, pressure and average
particle number respectively. Other thermodynamic quantities such as
internal energy $U$, Helmholtz free energy $F$, enthalpy $H$, Gibbs
function $G$, \textit{etc.}, can be calculated by the combination of
the quantities we obtained, based on their definitions or relations.

But for the QMDD model, $\Omega$ is not only a function of $T$, $V$,
$\mu$, but also depends explicitly on the quark density $\rho$
because of Eq.(7), $\Omega=\Omega(T,V,\mu,m^*(\rho))$. How to tackle
the thermodynamics self-consistently is still a problem. There have
been many wrangles in present references
\cite{Chakrabarty:1989,Chakrabarty:1991,Chakrabarty:1993,Benvenuto:1995a,
Benvenuto:1995b,Peng:2000,Zhang:2001,Peng:1999,Wen:2005,Wang:2000,
Zhang:2002,Zhang:2003, Wu:2005,Mao:2006}. The difficulty comes from
the first and second laws of reversible process thermodynamics
expressed by Eq.(8) and the partial derivatives by Eq.(9).
Obviously, some extra terms involving the derivatives of $m^*$ will
emerge when partial derivatives are calculated following Eq.(9).
Unfortunately, these extra terms for different treatments in
different references contradict each other. For example, for ideal
quark gas system of quasiparticle with effective quark mass
$m^*=m^*(\rho)$, the pressure and energy density $\varepsilon$ were
given by
\begin{eqnarray}
p&=&-\widetilde{\Omega}\equiv-\frac{\Omega}{V},\\
\varepsilon\equiv\frac{U}{V}&=&\widetilde{\Omega}+\sum_{i}\mu_{i}\rho_{i}-
T\frac{\partial\widetilde{\Omega}}{\partial T},
\end{eqnarray}
in Ref.\cite{Chakrabarty:1989,Chakrabarty:1991,Chakrabarty:1993}; or
given by
\begin{eqnarray}
p&=&-\left(\frac{\partial(\widetilde{\Omega}/\rho)}{\partial(1/\rho)}\right)_{T,\{\mu_i\}}
=-\widetilde{\Omega}+\rho\left(\frac{\partial\widetilde{\Omega}}{\partial
\rho}\right)_{T,\{\mu_i\}}\\
\varepsilon&=&\widetilde{\Omega}-\rho\left(\frac{\partial\widetilde{\Omega}}{\partial
\rho}\right)_{T,\{\mu_i\}}+\sum_i\mu_i\rho_i-T\left(\frac{\partial\widetilde{\Omega}}{\partial
T}\right)_{\{\mu_i\},\rho}.
\end{eqnarray}
in Ref.\cite{Benvenuto:1995a,Benvenuto:1995b}; and given by
\begin{eqnarray}
p&=&-\widetilde{\Omega}+\rho\left(\frac{\partial\widetilde{\Omega}}{\partial
\rho}\right)_{T,\{\mu_{i}\}},\\
\varepsilon&=&\widetilde{\Omega}+\sum_{i}\mu_{i}\rho_{i}-T\left(\frac{\partial\Omega}{\partial
T}\right)_{\{\mu_{i}\},\rho},
\end{eqnarray}
in Ref.\cite{Peng:2000}.

For the quark mass density and temperature dependent (QMDTD) model
\cite{Zhang:2001,Peng:1999,Wen:2005,Zhang:2002,Zhang:2003,Wu:2005,
Mao:2006} with quark mass
\begin{equation}
m_{q}=\frac{B(T)}{3\rho_B},\qquad(q=u,\overline{u},d,\overline{d}),
\end{equation}\begin{equation}
m_{s,\overline{s}}=m_{s0}+\frac{B(T)}{3\rho_B},
\end{equation}
where
\begin{equation}
B(T)=B_{0}\left[1-\left(\frac{T}{T_{c}}\right)^2\right],
\end{equation}
$p$ and $\varepsilon$ read
\begin{equation}
p=-\widetilde{\Omega}-V\frac{\partial\widetilde{\Omega}}{\partial
V}+\rho\sum_{i} \frac{\partial\widetilde{\Omega}}{\partial
m_{i}}\frac{\partial m_{i}}{\partial \rho},
\end{equation}
\begin{equation}
\varepsilon=\widetilde{\Omega}-\sum_{i}\mu_{i}
\frac{\partial\widetilde{\Omega}}{\partial\mu_{i}}-
T\frac{\partial\widetilde{\Omega}}{\partial
T}-T\sum_{i}\frac{\partial\widetilde{\Omega}} {\partial
m_{i}}\frac{\partial m_{i}}{\partial T},
\end{equation}
in Ref.\cite{Wen:2005}, respectively. The ambiguity arises from the
variable $\rho$ in $m^*$, because it is not one of the
characteristic variables of thermodynamic potential.

This paper involves from an attempt to clear above ambiguity and
suggests a method to calculate the thermodynamic quantities from
partial derivatives self-consistently. In next section, we will
address the traditional thermodynamic treatment with partial
derivative for QMDD model transparently and emphasize that the
difficulty can not be overcome by usual method. In fact, the
thermodynamic inconsistency for an ideal quasiparticle system with
effective mass $m^*(\rho,T)$ to describe the quark gluon plasma or
gluon plasma had been pointed out by many authors previously
\cite{Bannur:dual,Bluhm:2005,Yang:1995}. To avoid this difficulty,
in Sec.III, we argue that we can calculate the thermodynamic
quantities from equilibrium state and show that if we choose the
quasiparticle effective mass $m^*$ as a new independent degree of
freedom, we can calculate the thermodynamic quantities along
reversible process by usual partial derivative self-consistently. In
Sec.IV, we will apply this new treatment to study the QMDD model. We
show that we can get rid of previous difficulties with partial
derivative and take clear, reasonable and self-consistent results in
the fresh, which coincide with the results in equilibrium state. We
will consider the vacuum correction on the pressure and compare our
result with that given by MIT bag model in Sec.V. We will show that
even though our results are in agreement with that given by
Ref.\cite{Benvenuto:1995a, Benvenuto:1995b}, but the mechanism and
the physical reasons are complete different. In
Ref.\cite{Benvenuto:1995a,Benvenuto:1995b}, the negative pressure
comes from the partial derivative terms of $m^*$. But in our
treatment, it comes from the physical vacuum. Our mechanism is just
the same as that of MIT bag model. The last section includes a
summary and discussion.

\section{Inconsistency of traditional thermodynamic treatments with partial derivative}

We first repeat the traditional treatment based on Eqs.(1) and (2).
It has been established in Ref.\cite{Benvenuto:1995a,
Benvenuto:1995b,Peng:2000} systematically. In Ref.\cite{Peng:2000},
they derived
\begin{equation}
p=-\left(\frac{\partial\Omega}{\partial
V}\right)_{T,\mu}=-\left(\frac{\partial(V\widetilde{\Omega})}{\partial
V}\right)_{T,\mu}=-\left(\frac{\partial(V\widetilde{\Omega}/\overline{N})}{\partial
(V/\overline{N})}\right)_{T,\mu}=-\left(\frac{\partial(\widetilde{\Omega}/\rho)}{\partial
(1/\rho)}\right)_{T,\mu}=-\widetilde{\Omega}+\rho\left(\frac{\partial\widetilde{\Omega}}{\partial
\rho}\right)_{T,\mu}.
\end{equation}
Without losing generality, from here on we will display the formulae
with only one component for simplicity and clarity. Eq.(21) is
different from that of the normal expression
$p=-\Omega/V=-\widetilde{\Omega}$ by the last additional term due to
the density-dependence of effective mass. But Eq.(21) seems not so
solid if we pay attention to the invariables. Noting that the
chemical potential $\mu$ is determined by the particle density
constraint $\rho=\overline{N}/V$, and the partial derivative of
Eq.(21) is with respect to $V$ but fixed $T$ and $\mu$, or
equivalently fixed $T$ and $\rho$, in this process $\overline{N}$
must be changed. It is not a constant and so the third equal mark in
Eq.(21) can not hold. The result in Eq.(21) is error.

To avoid this ambiguity, in Ref.\cite{Wen:2005}, in stead of the
thermodynamic potential $\Omega$, the authors introduced the
Helmholtz free energy $F$ to change the characteristic variable
$\mu$ to the average particle number $\overline{N}$
\begin{equation}
dF=-SdT-pdV+\mu d\overline{N}.
\end{equation}
By using $F=\Omega+\mu\overline{N}$, they got
\begin{equation}
p=-\left(\frac{\partial F}{\partial
V}\right)_{T,\overline{N}}=-\left(\frac{\partial\Omega}{\partial
V}\right)_{T,\overline{N}}-\overline{N}\frac{\partial\mu}{\partial
V},
\end{equation}
\begin{equation}
S=-\left(\frac{\partial F}{\partial
T}\right)_{V,\overline{N}}=-\left(\frac{\partial\Omega}{\partial
T}\right)_{V,\overline{N}}-\overline{N}\frac{\partial\mu}{\partial
V}.
\end{equation}
After some calculations, they gave
\begin{equation}
p=-\frac{\partial\Omega}{\partial
V}-\left(\overline{N}+\frac{\partial\Omega}
{\partial\mu}\right)\frac{\partial\mu}{\partial
V}+\frac{\rho}{V}\frac{\partial\Omega}{\partial m^*}\frac{\partial
m^*}{\partial\rho},
\end{equation}
\begin{equation}
S=-\frac{\partial\Omega}{\partial
T}-\left(\overline{N}+\frac{\partial\Omega}
{\partial\mu}\right)\frac{\partial\mu}{\partial
T}-\frac{\partial\Omega}{\partial m^*}\frac{\partial m^*}{\partial
T}.
\end{equation}
For the internal energy, they required
\begin{equation}
\overline{N}=-\left(\frac{\partial\Omega}{\partial\mu}\right)_{V,m^*,T},
\end{equation}
and found
\begin{equation}
U=\Omega+\mu\overline{N}+TS=\Omega+\mu\overline{N}-T\frac{\partial\Omega}{\partial
T}-T\frac{\partial\Omega}{\partial m^*}\frac{\partial m^*}{\partial
T}.
\end{equation}
But in their formulae, the invariable quantities for partial
derivative have not been written down explicitly. To show the errors
of their calculation, we write down these fixed quantities in the
following by using the standard mathematical chain rule of partial
derivative with composition function:
\begin{eqnarray}
p&=&-\left(\frac{\partial F}{\partial
V}\right)_{T,\overline{N}}=-\left(\frac{\partial\Omega}{\partial
V}\right)_{T,\overline{N}}-\overline{N}\left(\frac{\partial\mu}{\partial
V}\right)_{T,\overline{N}}\nonumber\\
&=&-\left(\frac{\partial\Omega}{\partial
V}\right)_{T,\overline{N},\mu,m^*}-\left(\frac{\partial\Omega}
{\partial\mu}\right)_{T,\overline{N},m^*,V}\left(\frac{\partial\mu}{\partial
V}\right)_{T,\overline{N}}-\left(\frac{\partial\Omega}{\partial
m^*}\right)_{T,\overline{N},\mu,V}\left(\frac{\partial m^*}{\partial
V}\right)_{T,\overline{N}}-\overline{N}\left(\frac{\partial\mu}{\partial
V}\right)_{T,\overline{N}}\nonumber\\
&=&-\left(\frac{\partial\Omega}{\partial
V}\right)_{T,\overline{N},\mu,m^*}-\left[\overline{N}+\left(\frac{\partial\Omega}
{\partial\mu}\right)_{T,\overline{N},m^*,V}\right]\left(\frac{\partial\mu}{\partial
V}\right)_{T,\overline{N}}+\frac{\rho}{V}\left(\frac{\partial\Omega}{\partial
m^*}\right)_{T,\overline{N},\mu,V}\left(\frac{\partial
m^*}{\partial\rho}\right)_{T,\overline{N}},
\end{eqnarray}
\begin{equation}
S=-\left(\frac{\partial\Omega}{\partial
T}\right)_{V,\overline{N},\mu,m^*}-\left[\overline{N}+\left(\frac{\partial\Omega}
{\partial\mu}\right)_{V,\overline{N},m^*,T}\right]\left(\frac{\partial\mu}{\partial
T}\right)_{V,\overline{N}}-\left(\frac{\partial\Omega}{\partial
m^*}\right)_{V,\overline{N},\mu,T}\left(\frac{\partial m^*}{\partial
T}\right)_{V,\overline{N}}.
\end{equation}
Eqs.(29) and (30) are just Eqs.(25) and (26) with the unchanged
variable explicitly written out. Note that many partial derivatives
are with four fixed variables. The term
$\left(\frac{\partial\Omega}{\partial
m^*}\right)_{V,\overline{N},\mu,T}$ must vanish because
$m^*=m^*(T,\rho)$, $\rho=\overline{N}/V$, to fix $V$, $\overline{N}$
and $T$ is essentially to fix $m^*$. It means all derivatives
involving $m^*$ in Eqs.(29) and (30) should be zero, therefore the
extra terms $\left(\frac{\partial\Omega}{\partial
m^*}\right)_{T,\overline{N},\mu,V}\left(\frac{\partial
m^*}{\partial\rho}\right)_{T,\overline{N}}$ and
$\left(\frac{\partial\Omega}{\partial
m^*}\right)_{V,\overline{N},\mu,T}\left(\frac{\partial m^*}{\partial
T}\right)_{V,\overline{N}}$ can not survive. In fact, according to
the definition of Gibbs function $G=\mu\overline{N}=U-Ts+pV$, we can
directly find $\Omega=-pV$. The extra terms involving derivative of
$m^*$ will surely destroy this relation.

The inconsistency dose not only appear in the expression of $p$ and
$S$, but also in $\overline{N}$. If we calculate the particle number
$\overline{N}$ from Eqs.(8) and (9), we obtain
\begin{eqnarray}
\overline{N}&=&\left(\frac{\partial\Omega(T,V,\mu,m^*)}{\partial
\mu}\right)_{T,V,m^*}-\left(\frac{\partial\Omega(T,V,\mu,m^*)}{\partial
m^*}\right)_{T,V,\mu}\left(\frac{\partial
m^*(T,\rho)}{\partial\mu}\right)_{T,V}\nonumber\\
&=&\sum_i\frac{g_i}{e^{\beta(\sqrt{m^{*2}+k_i^2}-\mu)}+1}-\left(\frac{\partial\Omega(T,V,\mu,m^*)}{\partial
m^*}\right)_{T,V,\mu}\left(\frac{\partial
m^*(T,\rho)}{\partial\mu}\right)_{T,V},
\end{eqnarray}
where the summation is over all the quantum states with degeneracy
$g_i$. Since $\rho=\overline{N}/V$ is usually taken as a constraint
to decide the value of $\mu$, $\left(\frac{\partial
m^*(T,\rho)}{\partial\mu}\right)_{T,V}$ dose not vanish if $m^*$
depends on $\rho$ explicitly, so the extra term in Eq.(31) modifies
the particle number of the system. For an ideal quasiparticle
system, the number of quasiparticle $\overline{N}=\sum_ig_in_i$,
which is just the first term of Eq.(31), then we come to a
conclusion that the result of Eq.(31) is incorrect. Besides, this
result in Eq.(31) also conflicts with the requirement in Eq.(27),
then the inconsistency of the traditional treatment with partial
derivative is transparent.

In fact, the inconsistency of usual thermodynamic treatment for
ideal quasiparticle system has been shown by many authors
\cite{Bannur:dual,Bluhm:2005,Yang:1995,Wang:2000}. This section is
not a new result. Our aim is to demonstrate the inconsistency for
thermodynamic treatment with partial derivative along a reversible
process explicitly.

\section{A thermodynamically consistent treatment}

In Sec.II, we have shown the inconsistency of traditional
thermodynamic treatment with partial derivative along reversible
process. In this section, we will suggest a consistent thermodynamic
treatment for quasiparticle system with density and/or temperature
dependent particle mass $m^*(\rho,T)$. We want to point out that our
method is universal and QMDD model is just a special example for
using this method. Our method can also be applied to discuss the
thermodynamic behavior of the quark mass density and temperature
dependent (QMDTD) model \cite{Zhang:2001,Zhang:2002,Zhang:2003,
Wu:2005,Mao:2006,Wen:2005,Peng:1999} or other ideal quasiparticle
system.

Noticing that at a fixed instant of reversible process, the system
is at an equilibrium state. Denote the temperature and density of
this system as $T_0$ and $\rho_0$, respectively, then the effective
mass of the quasiparticle becomes constant $m^*(T_0,\rho_0)=m_0$.
The system reduces to a usual ideal gas system with constant mass
$m_0$ quasiparticles. For this equilibrium state, the corresponding
thermodynamic quantities can be directly obtained. For example, for
the case of one component Fermi system:
\begin{eqnarray}
\overline{N}&\equiv&\rho V=\sum_ig_in_i=\sum_i\frac{g_i}{e^{\beta(\sqrt{m_0^2+k_i^2}-\mu)}+1},\\
\Omega&=&-\sum_ig_ikT\ln(1+e^{-\beta(\sqrt{m_0^2+k_i^2}-\mu)}),\\
U&=&\sum_ig_in_i\epsilon_i=\sum_i\frac{g_i\sqrt{m_0^2+k_i^2}}{e^{\beta(\sqrt{m_0^2+k_i^2}-\mu)}+1},\\
G&=&\overline{N}\mu=\sum_ig_in_i\mu=\sum_i\frac{g_i\mu}{e^{\beta(\sqrt{m_0^2+k_i^2}-\mu)}+1},\\
S&=&\frac{U-\Omega-G}{T},\\
p&=&-\frac{\Omega}{V},
\end{eqnarray}
where $n_i$ is the particle number of the $i$th state and $g_i$ is
the corresponding degeneracy.

It can be clearly seen that these formulae are the same as those of
the standard ideal gas. In fact, for example, if we derive the
energy from the very beginning of the statistical definition in
grand canonical distribution, we have
\begin{eqnarray}
U=\overline{E}&=&\frac{1}{\Xi}\sum_NE_Ne^{-\beta(E_N-\mu N)}\nonumber\\
&=&\frac{1}{\Xi}\sum_N\left[\left(-\frac{\partial}{\partial\beta}\right)_{\mu,m^*,V}+\mu
N\right]e^{-\beta(E_N-\mu N)}\nonumber\\
&=&-\left(\frac{\partial\ln\Xi}{\partial\beta}\right)_{\mu,m^*,V}+\mu\overline{N},
\end{eqnarray}
where the invariables of $\mu$ is clearly seen, while $m^*$ and $V$
are fixed since $E_N$ is a function of $m^*$ and $V$ for each $N$.
Take the quasiparticle ideal gas thermodynamic potential, Eq.(33),
which means the grant partition function satisfies
\begin{equation}
\ln\Xi=\sum_ig_i\ln(1+ e^{-\beta(\sqrt{m^{*2}+k_i^2}-\mu)}),
\end{equation}
it is easy to find
\begin{eqnarray}
U&=&-\left(\frac{\partial\ln\Xi}{\partial\beta}\right)_{\mu,m^*,V}+\mu\overline{N}\nonumber\\
&=&\sum_ig_in_i\left(\sqrt{m^{*2}+k_i^2}-\mu\right)+\mu\overline{N}\nonumber\\
&=&\sum_ig_in_i\epsilon(m^*,k_i),
\end{eqnarray}
where $n_i=\frac{1}{e^{\beta[\epsilon(m^*,k_i)-\mu]}+1}$ is the
average particle number of the ith state. This result consists with
the interaction-free quasiparticle picture of the QMDD model.

We see, from Eqs.(32)-(37), the contribution of medium effect is
included in the effective value of mass, and appears in the
exponential of the Fermi distribution. A remarkable property of
these formulae is that the extra terms related to the partial
derivative of $m^*$ do not appear. This is reasonable because these
thermodynamic quantities are functions of equilibrium state, they do
not depend on the change of the quasiparticle mass, since at this
state the quasiparticle mass $m^*(T,\rho)$ becomes constant
$m_0(T_0,\rho_0)$.

To compare above treatment with those given in Sec.II, as an
example, we calculate the entropy of the ideal quasiparticle system.
Denote the entropy calculated by equilibrium state as $S_{sta}$.
From Eq.(34)-(37), we have:
\begin{eqnarray}
S_{sta}&=&\frac{U-G-\Omega}{T}\nonumber\\
&=&\sum_ig_i\left[\frac{n_i\sqrt{m_0^2+k_i^2}-n_i\mu}{T}
+k\ln(1+e^{-\beta(\sqrt{m_0^2+k_i^2}-\mu)})\right]\nonumber\\
&=&k\sum_ig_i\left[n_i\beta(\sqrt{m_0^2+k_i^2}-\mu)
+\ln\left(\frac{e^{\beta(\sqrt{m_0^2+k_i^2}-\mu)}+1}
{e^{\beta(\sqrt{m_0^2+k_i^2}-\mu)}}\right)\right]\nonumber\\
&=&k\sum_ig_i\left[n_i\ln\left(\frac{1}{n_i}-1\right)+\ln\left(\frac{1}{1-n_i}\right)\right]\nonumber\\
&=&-k\sum_ig_i[n_i\ln(n_i)+(1-n_i)\ln(1-n_i)].
\end{eqnarray}
This is a familiar formula for equilibrium state whose physical
meaning is transparent. While we denote the entropy calculated by
the partial derivative of $\Omega$ following Eqs.(8) and (9) as
$S_{der}$, we have
\begin{eqnarray}
S_{der}&=&-\left(\frac{\partial\Omega(T,V,\mu,m^*(T,\rho))}{\partial
T}\right)_{V,\mu}\nonumber\\
&=&-\left(\frac{\partial\Omega(T,V,\mu,m^*)}{\partial
T}\right)_{V,\mu,m^*}-\left(\frac{\partial\Omega(T,V,\mu,m^*)}{\partial
m^*}\right)_{T,V,\mu}\left(\frac{\partial m^*(T,\rho)}{\partial
T}\right)_{V,\mu},
\end{eqnarray}
and for the equilibrium state with $m^*(T_0,\rho_0)=m_0$,
\begin{eqnarray}
&&-\left(\frac{\partial\Omega(T,V,\mu,m^*)}{\partial
T}\right)_{V,\mu,m^*=m_0}\nonumber\\
&=&k\sum_ig_i\ln\left(1+e^{(\mu-\sqrt{m_0^2+k_i^2})/kT}\right)+kT\sum_ig_i\frac{e^{(\mu-\sqrt{m_0^2+k_i^2})/kT}
\left(\frac{\sqrt{m_0^2+k_i^2}-\mu}{kT^2}\right)}{1+e^{(\mu-\sqrt{m_0^2+k_i^2})/kT}}\nonumber\\
&=&k\sum_ig_i\ln\left(\frac{1}{1-n_{i}}\right)+k\sum_ig_in_{i}\ln\left(\frac{1}{n_{i}}-1\right)\nonumber\\
&=&-k\sum_ig_i[n_{i}\ln n_{i}+(1-n_{i})\ln(1-n_{i})]=S_{sta}.
\end{eqnarray}

The first term of Eq.(42) is just the result given by equilibrium
state. The difference between these two treatments is significant.
They can not be accorded together. Noticing that the contribution of
medium effect at equilibrium state is included within the value of
$m^*$ in the distribution, and the entropy describing disorder of
quasiparticles in a system dose not depend on the intrinsic quantity
such as effective mass of quasiparticle, the correctness of
$S_{sta}$ is obvious. The thermodynamic consistency of the new
treatment is transparent because it is based on an equilibrium state
and $m_0(T_0,\rho_0)$ is constant.

If one hopes to extend this treatment to a reversible process, the
accordance between the results from calculation along a reversible
process and results obtained at a fixed equilibrium state must hold.
So in order to get a self-consistent calculation between equilibrium
state and reversible process in thermodynamics, we are compelled by
above discussions to introduce an intrinsic degree of freedom $m^*$
for quasiparticle in thermodynamic system to describe medium effect.
We should rewrite Eq.(8) as:
\begin{equation}
d\Omega=-SdT-pdV-\overline{N}d\mu+Xdm^*,
\end{equation}
then Eq.(9) becomes
\begin{equation}
S=-\left(\frac{\partial \Omega}{\partial T}\right)_{V,\mu,m^*},\quad
p=-\left(\frac{\partial \Omega}{\partial V}\right)_{T,\mu,m^*},\quad
\overline{N}=-\left(\frac{\partial \Omega}{\partial
\mu}\right)_{T,V,m^*},\quad X=\left(\frac{\partial\Omega}{\partial
m^*}\right)_{T,V,\mu},
\end{equation}
where $X$ is an extensive quantity corresponding to the intensive
variable $m^*$. In Eq.(44), the intrinsic degree of freedom $m^*$
for quasiparticle has been added as an independent variable in
thermodynamic system. The quantities $S$, $p$ and $\overline{N}$
shown in Eq.(45) are in agreement with the results obtained by the
formulae of equilibrium state in Eqs.(32)-(37), because we have
fixed $m^*$ as an unchanged parameter in the partial derivative
calculations. For example, the pressure given by Eq.(9) is
\begin{eqnarray}
p&=&-\left(\frac{\partial\Omega(T,V,\mu,m^*(\rho_B))}{\partial
V}\right)_{T,\mu} \nonumber\\
&=&-\left(\frac{\partial\Omega(T,V,\mu,m^*)} {\partial
V}\right)_{T,\mu,m^*}-\left(\frac{\partial\Omega(T,V,\mu,m^*)}{\partial
m^*}\right)_{T,V,\mu}\left(\frac{\partial
m^*(\rho_B)}{\partial V}\right)_{T,\mu} \nonumber\\
&=&-\frac{\Omega}{V}-\left(\frac{\partial\Omega(T,V,\mu,m^*)}{\partial
m^*}\right)_{T,V,\mu}\left(\frac{\partial m^*(\rho_B)}{\partial
V}\right)_{T,\mu},
\end{eqnarray}
while given by Eq.(45) as
\begin{equation}
p=-\left(\frac{\partial\Omega(T,V,\mu,m^*)}{\partial
V}\right)_{T,\mu,m^*}=-\frac{\Omega}{V}.
\end{equation}
In Eq.(46), the first term is just the result of equilibrium state,
but the second term, according to the treatment in Sec.II, will lead
to inconsistency in thermodynamics.

Usually the thermodynamic parameters such as $S$, $p$, $T$... depend
on the whole system. They are independent with the intrinsic
property of the particle or subsystem, no matter the subsystem is a
simple point particle or a quasiparticle with inner structure and
different intrinsic properties. Ordinary thermodynamic variables
depend on the collection of the subsystem only. Similarly, the mass
is an intrinsic quantity of a particle, it does not affect on
collective thermodynamic properties of the whole system. When we
consider the medium effect or the confinement mechanism, and
summarize this effect into the effective mass $m^*(T,\rho)$ under
quasiparticle approximation, the dynamic interaction can be
concentrated on the effective mass $m^*$ of quasiparticle by using
the finite temperature quantum field calculation \cite{Das:1997} or
directly by confinement \textit{anst\"{a}ze}. But the macro
thermodynamic variables cannot describe these micro dynamic
interactions. We must choose new variables to represent these
dynamic interactions or the medium effect. Obviously, the effective
mass $m^*$ appears as the suitable independent variable. Further
more, we want to emphasize that the application of this new
independent variable $m^*$ to rewrite Eqs.(8) and (9) is limited to
quasiparticle approximation only. In quasiparticle approximation,
the interaction between particles, the medium effect or the
confinement mechanism are summarized in the effective mass $m^*$.
Then $m^*$ becomes an independent variable to represent all those
physical effects, its dependence on density and temperature is due
to equilibrium condition of the whole system. Introducing $m^*$ in
quasiparticle physical picture to represent the medium effect and
taking it as a variable is a twin in thermodynamics of quasiparticle
system. If we employ finite temperature quantum field theory to
calculate the thermodynamic potential beyond quasiparticle
approximation \cite{Blaizot:2001,Zhang:1997,Shiomi:1994}, we can use
Eq.(2) to calculate the thermodynamic quantities and investigate the
interaction between particles by different orders of Feynman
diagrams. Then we will not need to introduce the quasiparticle and
use the corresponding Eqs.(44) and (45).

\section{QMDD Model}

To illustrate the basic difference between our treatment based on
Eqs.(44) and (45) and those of traditional treatment based on
Eqs.(8) and (9), we employ QMDD model as an example to calculate the
thermodynamic property of strange quark matter. We fix the
parameters in Eqs.(1) and (2) $B=170$ MeVfm$^{-3}$, $m_{s0}=150$
MeV, as in Ref.\cite{Zhang:2002}. The temperature is set at $T=0$,
as that of Ref.\cite{Chakrabarty:1991,Benvenuto:1995a,Peng:2000},
for convenience of comparison. Our results are shown in Fig.1 and
Fig.2. In Fig.1, we draw the curves of energy per baryon
$\varepsilon/\rho_B$ vs. baryon number density $\rho_B$, where the
solid curve refers to our treatment and the dashed curve refers to
treatment in Ref.\cite{Benvenuto:1995a}. We see that the saturation
point of the solid curve is $\varepsilon/\rho_B=906.3$ MeV at
$\rho_B=0.433$ fm$^{-3}$, which locates in a reasonable range. The
saturation point of the dashed curve is $\varepsilon/\rho_B=1042.8$
MeV at $\rho_B=0.692$ fm$^{-3}$. It is higher than our result
because it has an additional term
$\rho_{B}\left(\frac{\partial\widetilde{\Omega}}{\partial
\rho_{B}}\right)_{T,\{\mu_{i}\}}$.

The equation of state for different treatments are shown in Fig.2,
where the solid curve refers to our treatment, and the dotted curve,
dashed curve refers to that given by Ref.\cite{Benvenuto:1995a} and
Ref.\cite{Peng:2000}, respectively. We see from Fig.2 that there are
two remarkable differences between the solid curve and others.
Firstly, the pressure is always positive in our treatment. This is
of course reasonable because our Hamiltonian
$H=\sum_{k,s}\varepsilon_{ks}a^+_{ks}a_{ks}$ is an ideal gas
Hamiltonian, it cannot give negative pressure in principle. But for
the dashed curve and dotted curve, the pressure becomes negative in
the small energy density regions due to the incorrect modification
term
\begin{equation}
\rho_{B}\left(\frac{\partial\widetilde{\Omega}}{\partial
\rho_{B}}\right)_{T,\{\mu_{i}\}}=-\sum_i\frac{g_i}{48\pi^2}\frac{4m^*_iB}{\rho_B}
\left[\mu_i\sqrt{\mu_i^2-m_i^{*2}}-m_i^{*2}
\ln\left(\frac{\mu_i+\sqrt{\mu_i^2-m_i^{*2}}}{m_i^*}\right)\right].
\end{equation}
Secondly, although the tendencies of all the curves are similar in
large energy density region, in small energy density region, their
behavior are very different. In particular, the end points at
$\rho_B\rightarrow0$ for these three curves are different. They are
$\{\varepsilon=170 \textrm{MeV fm}^{-3}, p=0\}$ for solid curve,
$\{\varepsilon=340 \textrm{MeV fm}^{-3}, p=-170 \textrm{MeV
fm}^{-3}\}$ for dashed curve and $\{\varepsilon=170 \textrm{MeV
fm}^{-3}, p=-170 \textrm{MeV fm}^{-3}\}$ for dotted curve,
respectively.

Can we get negative pressure pressure in QMDD model, as that of the
MIT bag model? We will answer this question in next section.

\section{The contribution of physical vacuum}

To answer this question, let us recall that in the MIT bag model the
negative pressure comes from the physical vacuum. If we extend the
quasiparticle Hamiltonian to
\begin{equation}
H=\sum_{k,s}\varepsilon_{ks}a^+_{ks}a_{ks}+H_0,
\end{equation}
where $H_0$ is the system energy in the absence of quasiparticle
excitations. Noting that in a quasiparticle system, the physical
vacuum energy $H_0$ is possibly a function of $T$ and $\rho$ since
$\varepsilon_{ks}$ depends on $T$ and $\rho$ via effective mass
$m^*$. Corresponding to $H_0$, the additional term of thermodynamic
potential is $\Omega_0$, and the pressure becomes
\begin{equation}
p'=-\frac{\Omega+\Omega_0}{V}=p-B_0,
\end{equation}
the energy density becomes $\varepsilon+B_0$, but the entropy
density $s=(\varepsilon+p-\sum_i\mu_i\rho_i)/T$ is unchanged.

Now we turn to determine $\Omega_0$ in QMDD model. By means of the
vacuum correction to guarantee the thermodynamic consistency of
quasiparticle system, it was first suggested by Gorenstein and Yang
\cite{Yang:1995}, they gave a set of constraints. While in
Ref.\cite{Wang:2000}, $\Omega_0$ is chosen to satisfy
\begin{equation}
\left(\frac{\partial(\widetilde{\Omega}+\widetilde{\Omega}_0)}{\partial\rho_B}\right)_{\{\mu_i\}}=0,
\end{equation}
where $\widetilde{\Omega}_0=\Omega_0/V$ is virtually $B_0$ in
Eq.(50). At zero temperature, Eq.(51) reduces to
\begin{equation}
d\widetilde{\Omega}_0(\rho_B)=-\left(\frac{\partial\widetilde{\Omega}}{\partial
m^*}\right)_{T=0,V,\mu}\frac{dm^*}{d\rho_B}d\rho_B,
\end{equation}
because $m^*$ depends on $\rho_B$ only, $\widetilde{\Omega}_0$ can
be obtained by integration. Based on Eqs.(1), (2) and (33), we
obtain
\begin{eqnarray}
\frac{d\widetilde{\Omega}_0(\rho_B)}{d\rho_B}
&=&-\left(\frac{\partial\widetilde{\Omega}}{\partial\rho_B}\right)_{T=0,\{\mu_i\}}\nonumber\\
&=&\sum_i\frac{g_iBm_i^*}{12\pi^2\rho_B^2}\left[\mu_i\sqrt{\mu_i^2-m^{*2}_i}
-m^{*2}_i\ln\left(\frac{\mu_i+\sqrt{\mu_i^2-m^{*2}_i}}{m^*_i}\right)\right],
\end{eqnarray}
therefore
\begin{equation}
\widetilde{\Omega}_0(\rho_B)=\int_{\rho_0}^{\rho_B}\sum_i\frac{g_iBm_i^*}{12\pi^2\rho_B^2}\left[\mu_i\sqrt{\mu_i^2-m^{*2}_i}
-m^{*2}_i\ln\left(\frac{\mu_i+\sqrt{\mu_i^2-m^{*2}_i}}{m^*_i}\right)\right]d\rho_B+\widetilde{\Omega}_0(\rho_0).
\end{equation}
Chosen $\widetilde{\Omega}_0(\infty)=B_\infty$, as that of
Ref.\cite{Wang:2000}, we get
\begin{equation}
\widetilde{\Omega}_0(\rho_B)=-\int_{\rho_B}^{\infty}\sum_i\frac{g_iBm_i^*}{12\pi^2\rho_B^2}\left[\mu_i\sqrt{\mu_i^2-m^{*2}_i}
-m^{*2}_i\ln\left(\frac{\mu_i+\sqrt{\mu_i^2-m^{*2}_i}}{m^*_i}\right)\right]d\rho_B+B_\infty.
\end{equation}
Eqs.(53)-(55) are just Eqs.(7),(8) and (14) in Ref.\cite{Wang:2000},
then the stability and the negative pressure have been shown there.
The negative pressure and the stable point dose exist, we do not
recalculate the numerical results here again.

Of course, the constraint Eq.(51) is one of the possible ways to get
$\widetilde{\Omega}_0$ only. The aim of this choice is to
demonstrate that we can get negative pressure and a stable strange
quark matter. We must emphasize here again that the negative
pressure as well as the stability of the strange quark matter come
from physical vacuum. It is completely different from that of
Ref.\cite{Benvenuto:1995a,Benvenuto:1995b,Peng:2000}, in which the
negative pressure comes from the extra partial derivative term
$\rho_B\frac{\partial\Omega}{\partial\rho_B}$. As has been shown in
Sec.II, this term cannot exist since it destroy the thermodynamic
consistency. Our mechanism is the same as that of MIT bag model. The
conclusion \cite{Benvenuto:1995a} that the properties of strange
quark matter in QMDD model are nearly the same as those obtained in
the MIT bag model is still valid, but their argument was incorrect.
The correct argument is shown by Sec.III-V.

\section{Summary and discussion}

In summary, we have shown the shortcomings of the previous
treatments for QMDD model, which are based on the partial derivative
of thermodynamical functions along reversible process. A new method
is suggested. For the reversible process, we have introduced a new
intrinsic degree of freedom $m^*$ for the system of quasiparticle.
We have proved that the thermodynamic quantities calculated by the
partial derivatives concerning this independent variable are in
agreement with those obtained by the equilibrium state. The
difficulties and controversies in previous references are removed.
We also find that the properties of strange quark matter in QMDD
model are nearly the same as those obtained in MIT bag model
\cite{Wu:2005}, if we take the vacuum contribution into account. Our
method is applicable if the quasiparticle with effective mass is
introduce to represent the medium effect, since such quasiparticle
models are quite commonly used in nuclear matter, quark matter and
quark-gluon plasma, we hope our discussion might help the
theoretical study in these fields.

\section*{Acknowledgements}
This work is supported in part by NNSF of China.

\begin{figure}[tbp]
\includegraphics[totalheight=16cm, width=16cm]{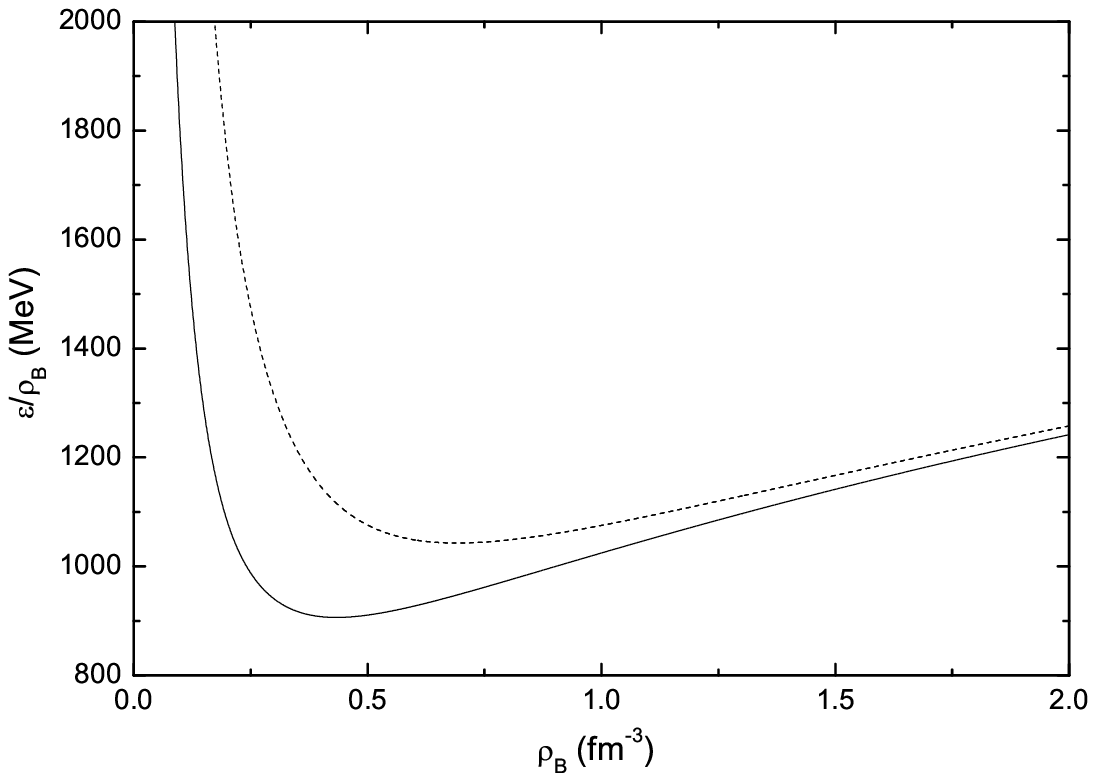}
\caption{Energy per baryon as a function of the baryon density
$\rho_{B}$ at $T=0$ for different treatments of the QMDD model. The
solid curve represents our treatment and the dashed line refers to
the treatment in Ref.\cite{Benvenuto:1995a}.} \label{fig1}
\end{figure}

\begin{figure}[tbp]
\includegraphics[totalheight=16cm, width=16cm]{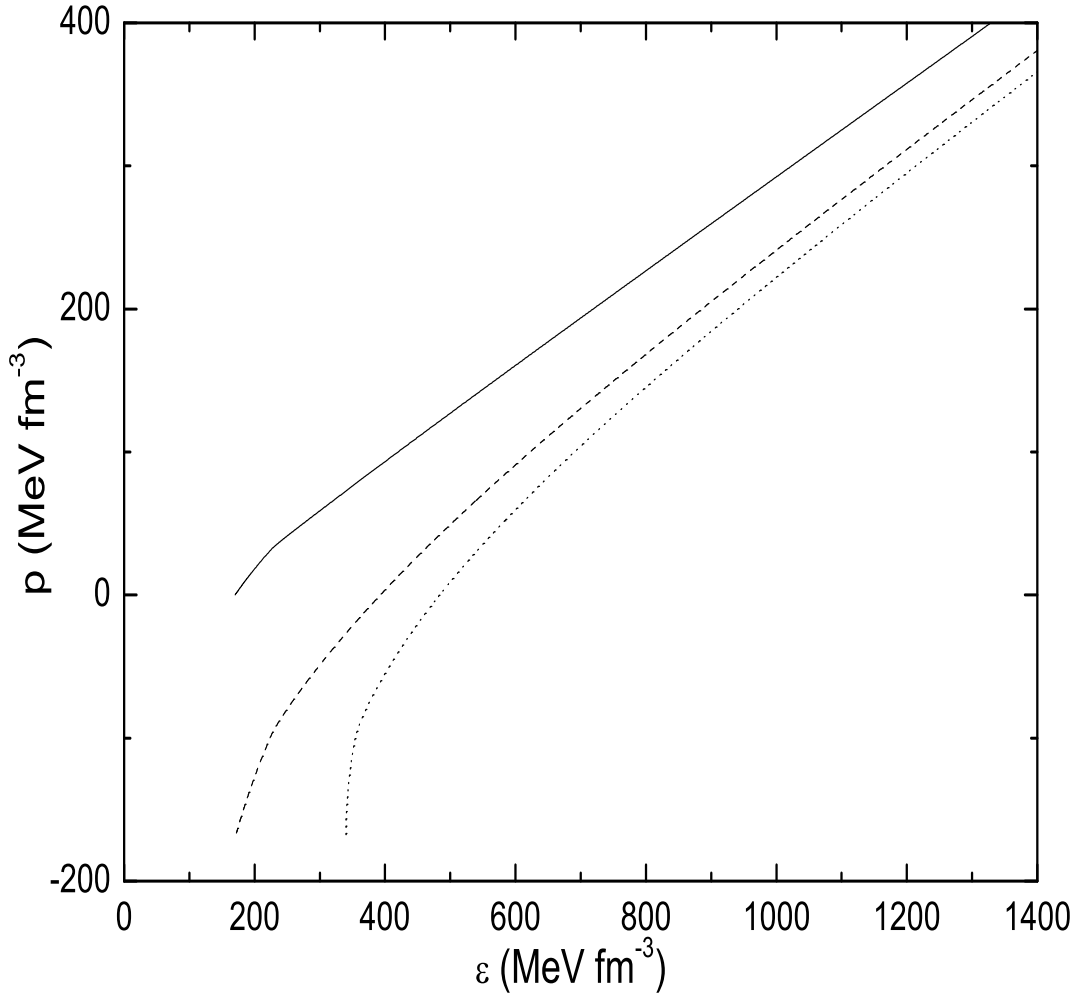}
\caption{Pressure $P$ as a function of the energy density
$\varepsilon=U/V$ for different treatments of the QMDD model. The
solid curve, dotted curve and dashed curve represent our treatment,
the treatment in Ref.\cite{Benvenuto:1995a} and the treatment in
Ref.\cite{Peng:2000}, respectively. The tendencies of these curves
are similar at large energy density region, but at small energy
density region, different treatments have quite different behaviors.
In our treatment, the pressure never goes to negative.} \label{fig2}
\end{figure}

\end{document}